\documentclass[12pt]{JHEP3}

\preprint{  }

\usepackage{epsfig,multicol}
\usepackage{amsmath}
\usepackage{graphics}
\usepackage{graphicx}
\usepackage{dcolumn}
\usepackage{amssymb}
\usepackage{amsthm}
\usepackage{amsfonts}
\usepackage{subfigure}
\usepackage{setspace}

\title{Distinguishing de Sitter universe from thermal Minkowski spacetime by Casimir-Polder-like force}
\author{Zehua Tian and
Jiliang  Jing\footnote{Corresponding author. Email:
jljing@hunnu.edu.cn}$^{1}$
\\ Department of Physics, and Key Laboratory of Low Dimensional Quantum Structures and Quantum Control of Ministry of Education, Hunan Normal University, Changsha, Hunan 410081, P. R. China}

\abstract{We demonstrate that the static ground state atom, which interacts with a conformally coupled massless scalar field in the de Sitter invariant vacuum, can obtain a position-dependent energy-level shift and this shift could cause a Casimir-Polder-like force on it. Interestingly no such force arises on the inertial atom bathed in a thermal radiation in the Minkowski universe. Thus, although the energy-level shifts of the static atom for these two cases are structurally the same, whether the energy-level shift causes the Casimir-Polder-like force,  in principle, could be as an indicator to distinguish de Sitter universe from the thermal Minkowski spacetime.}


\begin{document}

\section{Introduction}

The Casimir effect \cite{Casimir1}, deriving from the fluctuations of quantum vacuum due to the Heisenberg uncertainty principle, is related to many fundamental physical fields, such as the condensed matter physics, statistical physics, atomic physics, nanotechnology, even elementary particle physics, quantum field theory, gravitation and cosmology \cite{Klimchitskaya}. As one of its specific embodiments, Casimir-Polder force has been fruitfully applied to theoretical and experimental investigations \cite{Klimchitskaya}, like the display of the nonlocal properties of field correlations \cite{Rizzuto1}, probe of the entanglement \cite{Cirone} and detection of the Unruh effect \cite{Rizzuto2}. It is believed that the studies of the Casimir-Polder force, besides giving a striking manifestation of the existence of zero-point fluctuations, could also provide a deeper understanding of some novel phenomenons, e.g., dispersion forces, in real materials and provide some guidance as to how
to apply the Lifshitz theory to interpret the measurement results.

Because of the presence of conducting boundary, the vacuum fields modes will be reflected at which, and as a results of that, the vacuum fluctuations are sure to be modified correspondingly. Thus, as a respond of such modifications, a neutral electric polarizable atom, which keeps a distance far away from the conducting plate, is naturally expected to obtain a position-dependent energy-level shift due to the direct interaction between it and the scattered fields. Interestingly this shift, that can be modified by the motions of atom \cite{Audretsch,Rizzuto3}, different boundaries conditions \cite{Rizzuto3} and even curvature of spacetime \cite{zhou1}, may lead to an observable quantity, Casimir-Polder force between the neutral electric polarizable atom and the conducting plate. So, in this point, we can think that the reshaping of vacuum fluctuation, being a result of the reflection of field modes at boundary, causes the Casimir-Polder force. Especially, when an atom is placed in a curved spacetime, such as a black hole or an expanding universe, the curvature of spacetime can also scatter the vacuum field modes. Thus, it is naturally to ask whether the scattered vacuum field modes by the curvature will also induce a energy-level shift of or a Casimir-Polder-like force on the atom placed in the curved spacetime. And the other way round, whether such quantum effects resulting from zero-point fluctuation can be used to understand the property of spacetime, or distinguish different spacetimes.

We use the open quantum system approach to study the energy-level shift of and the Casimir-Polder-like force on a two-level atom, which is coupled to a conformally coupled massless scalar field in the de Sitter invariant vacuum. The reason for our special attention to de Sitter spacetime stems from the fact that de Sitter space, which is the simplest nontrivial curved background, enjoys the same degree of symmetry as Minkowski space (ten Killing vectors), and may be the candidate of our universe in the far past and the far future suggested by our current observations and the theory of inflation. On the other hand, the thermal effects perceived by a static detector on the de Sitter invariant vacuum are identical to that in the case of a thermal ensemble of field particles in flat spacetime \cite{Gibbons,Birrell}, such as the response of a single particle detector \cite{Birrell,Deser,Galtsov}, the correction to Lamb shift \cite{zhou1} and geometric phase \cite{Tian} of a two-level atom. Thus, it is worthy to ask whether it is possible to distinguish de Sitter spacetime from the thermal Minkowski spacetime, i.e., which unverse, de Sitter spacetime and thermal Minkowski universe, is that the inhabitants are exactly in.

In this paper, we investigate the energy-level shift and Casimir-Polder-like force together with general relativity theory with the hope of detecting spacetime curvature and especially distinguishing two important and generally studied universes, de Sitter universe and thermal Minkowski universe, in theory. We firstly treat a two-level atom as an open quantum system, and derive its energy-level shift from the open quantum system approach. Then we compare the atomic energy-level shift and Casimir-Polder-like force caused by this shift when the atom is static in de Sitter spacetime with that in thermal Minkowski spacetime. And finally we distinguish these two universes by examining whether the Casimir-Polder-like force is caused by the energy-level shift.

\section{Distinguishing de Sitter universe from thermal Minkowski spacetime by Casimir-Polder-like force}\label{section 1}

We assume a combined system, consisting of a detector and external fluctuating vacuum field, to be initially prepared in a uncorrelated state. Without loss of generality, the total Hamiltonian of the complicated system can be taken as
\begin{eqnarray}\label{Hamiltonian}
H=H_s+H_\phi+H_I,
\end{eqnarray}
where $H_s$ and $H_\phi$ are the Hamiltonian of the detector and scalar field, and $H_I$ represents their interaction.
For simplicity, we take a two-level atom as the detector, with Hamiltonian $H_s=\frac{1}{2}\omega_0\sigma_z$, where $\omega_0$ is its energy-level spacing and $\sigma_z$ is the Pauli matrix. The Hamiltonian describing the interaction between the atom and scalar field is described by $H_I=\mu(\sigma_++\sigma_-)\phi(x(\tau))$, in which $\mu$ is the coupling constant,  $\sigma_+$ ($\sigma_-$) is the atomic rasing (lowering) operator, and $\phi(x)$ corresponds to the scalar field operator.

Initially, the total density matrix of the complete system, atom plus field, is assumed to be $\rho_{tot}=\rho(0)\otimes|-\rangle\langle-|$, where $\rho(0)$ denotes the initial state of the atom, and $|-\rangle$ represents the vacuum state of the field. It is worthy to note that this state is uncorrelated at the beginning. In the frame of the atom, the equation of the motion of the combined system is given by
\begin{eqnarray}\label{motion equation}
\frac{\partial\rho_{tot}(\tau)}{\partial\tau}=-i[H,\rho_{tot}(\tau)],
\end{eqnarray}
where $\tau$ is the proper time of the atom, and $\rho_{tot}(\tau)$ denotes the time-dependent density matrix of the atom and the field. By tracing over the field degrees of freedom, i.e., $Tr_\phi[\rho_{tot}(\tau)]$, repeating the same processes in Ref. \cite{Breuer, Doukas} we can obtain the dynamics of the atom in the limit of weak coupling, which can be written in the Lindblad form \cite{Lindblad,Benatti}
\begin{eqnarray}\label{Lindblad equation}
\frac{\partial\rho(\tau)}{\partial\tau}&=&-i[H_{eff},\rho(\tau)]+\mathcal{L}[\rho(\tau)]
\end{eqnarray}
with\begin{eqnarray}\label{Effective H}
&&H_{eff}=\frac{1}{2}\Omega\sigma_z=\frac{1}{2}\{\omega_0+\mu^2\mathrm{Im}(\Gamma_++\Gamma_-)\}\sigma_z,
\\
&&\mathcal{L}[\rho(\tau)]=\sum^3_{j=1}[2L_j\rho L^\dagger_j-L^\dagger_jL_j\rho-\rho L^\dagger_jL_j],
\end{eqnarray}
where
$\Gamma_\pm=\int^{\infty}_{0}e^{i\omega_0s}G^+(s\pm i\epsilon)ds$,
$L_1=\sqrt{\frac{\gamma_-}{2}}\sigma_-$, $L_2=\sqrt{\frac{\gamma_+}{2}}\sigma_+$, $L_3=\sqrt{\frac{\gamma_z}{2}}\sigma_z$,
$\gamma_\pm=2\mu^2\mathrm{Re}\Gamma_\pm$,
$\gamma_z=0$,
$G^+(x-x')=\langle0|\phi(x)\phi(x')|0\rangle$ is the field correlation function, and $s=\tau-\tau'$. It is interesting to note that the effective Hamiltonian, $H_{eff}$, should be understood from two important parts, one is $\frac{1}{2}\omega_0\sigma_z$ which results from the internal energy of the atom, another term given by
\begin{eqnarray}\label{Ls H}
H_{LS}&=&\frac{1}{2}\mu^2\mathrm{Im}(\Gamma_++\Gamma_-)\sigma_z
\end{eqnarray}
denotes the energy-level shift of the atom due to the interaction between it and the external field. Thus the energy-level shifts of the ground state and excited state are $\delta E_-=-\frac{1}{2}\mu^2\mathrm{Im}(\Gamma_++\Gamma_-)$
and $\delta E_+=\frac{1}{2}\mu^2\mathrm{Im}(\Gamma_++\Gamma_-)$, respectively. The relative energy-level shift (Lamb shift)
is then
\begin{eqnarray}\label{LM S}
\Delta=\mu^2\mathrm{Im}(\Gamma_++\Gamma_-).
\end{eqnarray}


In order to study how the curvature of de Sitter spacetime affects the energy-level shift of a static atom in Eq. (\ref{LM S}), we need the Wightman function for the conformally coupled massless scalar field. To suitably describe the static atom in this spacetime, it is convenient to take the static de Sitter metric in the form of
\begin{eqnarray}\label{static coordinate}
ds^2=\big(1-\frac{r^2}{\alpha^2}\big)dt^2-\big(1-\frac{r^2}{\alpha^2}\big)^{-1}dr^2
-r^2 d\Omega^2.
\end{eqnarray}
We can see from Eq. (\ref{static coordinate}) that this metric possesses an event horizon at $r=\alpha=\sqrt{3/\Lambda}$ with $\Lambda$ being the cosmological constant, and usually we call it the cosmological horizon. It is of interest to note that an
observer at rest in this static coordinates system has a proper acceleration
\begin{eqnarray}\label{proper acceleraion}
a=\frac{r}{\alpha^2}\big(1-\frac{r^2}{\alpha^2}\big)^{-1/2}.
\end{eqnarray}

When vacuum fluctuations are concerned in a curved spacetime, a delicate issue arising then is how to determine
the vacuum state of the quantum field. In this letter, we will choose the de Sitter invariant vacuum sate
because it preserves the de Sitter invariance and is considered to be a natural choice of vacuum state in
this spacetime \cite{Allen1}. By solving the field equation in the static coordinate system, a set of modes will be obtained \cite{Mishima,Polarski1,Polarski2,Nakayama}. Then the Wightman function of the conformally coupled
massless scalar field for two spacetime points where the static atom locates can be easily derived as \cite{Polarski, Tikhonenko}
\begin{eqnarray}\label{wightman2}
G^+_{ds}(x, x')=-\frac{1}{16\pi^2\kappa^2\sinh^2(\frac{\tau-\tau'}{2\kappa}-i\epsilon)}.
\end{eqnarray}
Here $\kappa=\sqrt{g_{00}}\alpha$ and $\tau=\sqrt{g_{00}}t$.

Combining Eqs. (\ref{Effective H}) and (\ref{wightman2}), after some calculations, we obtain
\begin{eqnarray}\label{EH1}
\nonumber
H_{eff}&=&\frac{1}{2}\{\omega_0+\mu^2\mathrm{Im}(\Gamma_++\Gamma_-)\}\sigma_z
\\
&=&\frac{1}{2}\bigg\{\omega_0+\frac{\mu^2}{4\pi^2}\int^\infty_0 d\omega
P\bigg(\frac{\omega}{\omega+\omega_0}-\frac{\omega}{\omega-\omega_0}\bigg)\nonumber
\\
&&\times\bigg(1+\frac{2}{e^{2\pi\kappa\omega}-1}\bigg)\bigg\}
\sigma_z
\end{eqnarray}
for the atom at rest in the static coordinate system. Then its relative energy-level shift is given by
\begin{eqnarray}\label{dsRES}
\Delta_{DS}=\Delta_0+\Delta_s
\end{eqnarray}
with
\begin{eqnarray}\nonumber
\Delta_0&=&\frac{\mu^2}{4\pi^2}\int^\infty_0d\omega
P\bigg(\frac{\omega}{\omega+\omega_0}-\frac{\omega}{\omega-\omega_0}\bigg),
\\ \nonumber
\Delta_s&=&\frac{\mu^2}{2\pi^2}\int^\infty_0d\omega
P\bigg(\frac{\omega}{\omega+\omega_0}-\frac{\omega}{\omega-\omega_0}\bigg) \frac{1}{e^{2\pi\kappa\omega}-1},
\end{eqnarray}
where $\Delta_0$ is the same with the energy-level correction to an inertial two-level atom in a free Minkowski spacetime, which results from the fluctuation of the vacuum field that the atom coupled to. Obviously, $\Delta_s$ is the correction to the energy-level of the atom resulting from the thermal effect with a thermal factor $(e^{2\pi\kappa\omega}-1)^{-1}$. It is interesting to note that this term is similar to the correction to that of an inertial atom immersed in a thermal bath in Minkowski spacetime at the temperature $T_s=1/2\pi\kappa$ (discussed below). Thus, we conclude that due to the interaction between the static atom and the massless scalar field in de Sitter spacetime the energy-level of the atom is revised compared with that of the static atom in a free Minkowski spacetime. Here we need to note that $T_s=\sqrt{T^2_f+T^2_U}=\sqrt{(\frac{1}{2\pi\alpha})^2+(\frac{a}{2\pi})^2}$, where $T_f=\frac{1}{2\pi\alpha}$ is the Gibbons-Hawking temperature, and $T_U=\frac{a}{2\pi}$ is the Unruh temperature that is associated with the proper acceleration of
the static atom. Thus, the shift $\Delta_s$ means that besides the Gibbons-Hawking effect \cite{Gibbons}, the static atom also is subjected to a Unruh effect. It is worthwhile to note the Lamb shift of the static atom in de Sitter spacetime we obtain by open quantum system approach is the same with that of Ref. \cite{zhou1} derived from the Dalibard-Dupont-Roc-Cohen-Tannoudji formalism \cite{Dalibard1,Dalibard2}.

For comparison, we also discuss the energy-level shift of a static atom immersed in a thermal bath at temperature $T=1/\beta$ in Minkowski spacetime. For such case, the corresponding Wightman function is given by
\begin{eqnarray}\label{thermal WF}
G^+_{\beta}(x,x')=-\frac{1}{4\pi^2}\sum^{\infty}_{m=-\infty}
\frac{1}{(t-t'+im\beta-i\varepsilon)^2}.
\end{eqnarray}
Substituting Eq. (\ref{thermal WF}) into Eq. (\ref{LM S}),  we can also easily obtain the relative energy-level shift of the static atom immersed in the thermal bath in Minkowski spacetime,
\begin{eqnarray}\label{thRES}
\Delta_{MK}=\Delta_0+\Delta_\beta
\end{eqnarray}
with
\begin{eqnarray}\nonumber
\Delta_\beta&=&\frac{\mu^2}{2\pi^2}\int^\infty_0d\omega
P\big(\frac{\omega}{\omega+\omega_0}-\frac{\omega}{\omega-\omega_0}\big) \frac{1}{e^{\beta\omega}-1}.
\end{eqnarray}
$\Delta_\beta$ is the pure correction to the energy level of the atom coming from the thermal effect of the external field, and $\frac{1}{e^{\beta\omega}-1}$ is the thermal factor. Comparing Eq. (\ref{dsRES}) with (\ref{thRES}), obviously, these two corrections are the same  if we take $T=1/\beta=1/2\pi\kappa$, which means that only by the means of the correction to the energy level of the atom, the observer can't tell which universe, the de Sitter unverse or the thermal Minkowski spacetime, he is exactly in. In this regard, let's note that it is impossible to distinguish these two universes by the response of a single particle detector \cite{Deser,Galtsov} and the correction to geometric phase of an atom \cite{Tian} too. Thus, if the inhabitants wish to know whether this perceived thermality is a result of thermal field in Minkowski spacetime or property of de Sitter spacetime, they must be more creative.


Because the static atoms in de Sitter spacetime have an inherent acceleration which is position-dependent, naturally the energy-level shift discussed above depends on the position of the static atom too. Due to that, this position-dependent energy-level shift is similar to a Casimir-Polder potential and will cause a Casimir-Polder-like force on the static atom. Such force may provide us an avenue to differentiate the above two universes. Next we will show the Casimir-Polder-like force on the static atom at ground state in de Sitter spacetime. To obtain this force, we focus on its energy-level shift, which is given by
\begin{eqnarray}\label{energy shift1}
\delta E_-=\delta E_0+\delta E_{r},
\end{eqnarray}
where $\delta E_0=-\frac{1}{2}\Delta_0$ and $\delta E_r=-\frac{1}{2}\Delta_s$.
Obviously, $\delta E_0$, which is just the Lamb shift of an inertial two-level ground state atom interacting with a massless scalar field in the free Minkowski vacuum, is logarithmically divergent, but this divergence can be removed by taking a cutoff on the upper limit of the integral introduced by Bethe \cite{Bethe,Welton}. After doing like that, we obtain
\begin{eqnarray}\label{deltaE0}
\delta E_0\approx\frac{\mu^2\omega_0}{4\pi^2}\ln(\frac{m}{\omega_0}).
\end{eqnarray}
Because it is impossible to obtain the analytical result of $\delta E_r$, we will approximately study it in the low- and high-temperature limits, i.e., far away from and near the cosmological horizon, which is given by
\begin{eqnarray} \label{potential}
\delta E_{r}
&\approx&\bigg\{
\begin{array}{c}
-\frac{\mu^2\omega_0}{2\pi^2}\big[\frac{\pi^2T^2_s}{6\omega^2_0}+\frac{\pi^4T^4_s}{15\omega^4_0}\big]
~~~~(\omega_0\gg T_s)
\\
\frac{\mu^2\omega_0}{2\pi^2}\ln(\frac{\omega_0}{T_s})
~~~~~~~~~~~~~~~~~~(\omega_0\ll T_s).
\end{array}
\end{eqnarray}

Due to that $T_s$ is distance-dependent introduced above, naturally the energy-level shift (\ref{potential}) gives rise to a force on the atom at ground state that can be calculated by taking the first derivative with respect to $r$, which is
\begin{eqnarray}\label{force1}
F&\approx&
\bigg\{
\begin{array}{c}
\frac{\mu^2\omega_0}{2\pi^2}\big[\frac{4\pi^4T^4_fr}{3\omega^2_0g^2_{00}}
+\frac{16\pi^6T^6_fr}{15\omega^4_0g^3_{00}}\big]
~~(\omega_0\gg T_s)
\\
\frac{\mu^2\omega_0}{2\pi^2}\frac{4\pi^2T^2_fr}{g_{00}}~~~~~~~~~~~
~~~~~~~~(\omega_0\ll T_s).
\end{array}
\end{eqnarray}
Near the cosmological horizon, $r=\alpha$, the Casimir-Polder-like force is repulsive and actually diverges due to $g_{00}\rightarrow0$. It is interesting to note that the classical force that is needed to keep the atom static from falling into the horizon also diverges. When $r\rightarrow0$, i.e, when the atom approaches to the central of the cosmology, the Casimir-Polder-like force then obviously vanishes. However, it always exists as long as $r\neq0$. To the contrary, for the inertial atom immersed in a thermal bath in Minkowski spacetime, the first derivative of its energy-level shift (\ref{thRES}) versus position, $r$, is zero, because its energy-level shift is a constant, i.e., which is independent of $r$. Thus, in theory,
although the static atom in de Sitter spacetime behaves as if it were the static atom in a free Minkowski spacetime, which is immersed in a thermal bath at the temperature that is a square root of the sum of the squared Gibbons-Hawking temperature and the squared Unruh temperature associated with the atomic inherent acceleration, a further analysis shows that the energy-level shift of the static atom in de Sitter spacetime is position-dependent and thus causes a Casimir-Polder-like force on the atom, that is quite different from the case that the atom is immersed in the thermal Minkowski spacetime at the same temperature, which experiences no such force. As a result of that, this distinct difference between them, in principle, could tell the observer that he is in de Sitter spacetime or in a thermal bath in the Minkowski spacetime.

Now we focus on the numerical analysis of (\ref{force1}). If we take the energy-level spacing of the two-level atom as $\omega_0\sim10^9\mathrm{Hz}$, the condition $\omega_0\gg T_s$ approximately means $10^{28}\gg1/\sqrt{1-\xi^2}$ with $\xi=r/\alpha$. Obviously, such condition are satisfied even though the atom locates quite near the cosmological horizon, such as
$r=0.999999999\alpha$, for which the condition, $10^{28}\gg1/\sqrt{1-\xi^2}$, approximately turns out to be $10^{28}\gg10^4$. In this regard, let's note that the approximate result of the Casimir-Polder-like force on the static ground state atom in de Sitter spacetime, under the condition $\omega_0\gg T_s$, can be suitably applied to describing almost all different cases that the atom is placed at different positions from the center of de Sitter universe, $r=0$, to the point quite near the horizon. We assume that the atom is near the horizon and take $1/\sqrt{1-\xi^2}\sim10^{25}$, then the Casimir-Polder-like force is estimated to be $\sim10^{-14}\mathrm{N}$. In this case the atom, to avoid falling into the horizon, must have the acceleration $a\sim10^{16}\mathrm{m/s^2}$, which is a very large value. Such analysis, therefore, is of theoretical interest.


\section{Conclusions} \label{section 4}

In summary, we firstly obtain the energy-level shift of a two-level atom by the open quantum system approach. Then we discuss the Casimir-Polder-like force caused by such shift when the atom is placed in de Sitter universe and in thermal Minkowski universe. Interestingly, although the static atom coupled to the fluctuating scalar field in the de Sitter invariant vacuum behaves the same as if it were in a thermal bath in a flat spacetime at the same temperature, the energy-level shift for the former case will cause a Casimir-Polder-like force due to that it is position-dependent. Therefore, this distinct difference, in principle, can be used to distinguish de Sitter universe from the thermal Minkowski spacetime.

\begin{acknowledgments}

This work was supported by the  National Natural Science Foundation
of China under Grant No. 11175065; the National Basic Research of China under Grant No. 2010CB833004; the SRFDP under
Grant No. 20114306110003; Hunan Provincial Innovation Foundation For Postgraduate under Grant No CX2012B202.

\end{acknowledgments}


\end{document}